\documentclass[12pt]{article}

\usepackage{graphicx}

\usepackage{cite}

\def\beq{\begin{eqnarray}}
\def\eeq{\end{eqnarray}}

\begin{document}

\title{Comparison of alternative improved perturbative methods for nonlinear
oscillations }
\author{Paolo Amore and Alfredo Raya \\
Facultad de Ciencias, Universidad de Colima, \\
Bernal D\'{\i}az del Castillo 340, Colima, Colima, Mexico \and Francisco M.
Fern\'{a}ndez \\
INIFTA (Conicet,UNLP), \\
Diag. 113 y 64 S/N, Sucursal 4, Casilla de Correo 16, \\
1900 La Plata, Argentina}
\maketitle

\begin{abstract}
We discuss and compare two alternative perturbation approaches for the
calculation of the period of nonlinear systems based on the
Lindstedt--Poincar\'{e} technique. As illustrative examples we choose
one--dimensional anharmonic oscillators and the Van der Pol equation. Our
results show that each approach is better for just one type of model
considered here.
\end{abstract}

\section{\label{sec:Intro} Introduction}

There are several methods for removing the secular terms produced by the
straightforward application of perturbation theory to nonlinear motion \cite
{N81}. One of them is the Lindstedt--Poincar\'{e} technique (LPT) \cite
{N81,F00}, recently improved by Amore \emph{et al.} \cite
{AA03,AL04,AA03b,AM04} by means of the delta expansion and the principle of
minimal sensitivity (PMS) \cite{S81} giving rise to the LPLDE method.

There is also another approach that resembles the LPT \cite{M70} that we
will call alternative Lindstedt--Poincar\'{e} technique (ALPT) from now on.
The purpose of this letter is the discussion of those alternative approaches
and comparison of their results for simple nontrivial models.

In Sec.~\ref{sec:LP} we briefly review the LPT \cite{N81,F00} and
the LPLDE \cite{AA03,AL04,AA03b,AM04}. In Sec.~\ref{sec:ALP} we
outline the ALPT \cite {M70}. In all those cases we choose the
Duffing oscillator \cite{N81} as an illustrative example. In Sec.
\ref{sec:Other_AO} we apply the LPLDE and the ALPT to other
anharmonic oscillators. In Sec. \ref{sec:VDP} we apply both
methods to the Van der Pol equation. Finally, in
Sec.~\ref{sec:Results} we compare the frequencies of the motion
for the above mentioned models calculated with both approaches.

\section{\label{sec:LP} The Lindstedt-Poincar\'{e} Technique}

The LPT consists of the simultaneous expansion of the trajectory and
oscillation frequency in powers of the perturbation parameter. For
concreteness we choose the Duffing anharmonic oscillator as an illustrative
example, \cite{N81}:
\begin{equation}
\frac{d^{2}x(t)}{dt^{2}}+x(t)=-\mu x^{3}(t)\ ,  \label{eq:Duffing}
\end{equation}
with the initial conditions $x(0)=1$ and $\frac{dx(0)}{dt}=0$. The parameter
$\mu $ is a measure of the nonlinearity of the motion.

Straightforward application of perturbation theory to Eq. (\ref{eq:Duffing})
produces an approximate solution in the form of a $\mu $--power series $%
x(t)=\sum_{n=0}^{\infty }x_{n}(t)\mu ^{n}$ \cite{N81,F00}. It is well--known
that the equations for the coefficients $x_{n}(t)$ have resonant terms that
after integration give rise to nonperiodic contributions \cite{N81,F00}.
Such terms are unbounded in spite of the fact that the solution $x(t)$ is
known to be periodic for all $\mu >-1$. For this reason, the perturbative
solution fails at large time scales.

There are several approaches that provide approximate solutions free from
secular terms \cite{N81}. Here we are interested in the
Lindstedt--Poincar\'{e} technique that consists of introducing the actual
frequency of motion $\Omega $ into the equation of motion by means of a
time dilatation $\tau =\Omega t$ \cite{N81,F00}. Thus the Duffing equation (%
\ref{eq:Duffing}) becomes
\begin{equation}
\Omega ^{2}\ddot{X}(\tau )+X(\tau )=-\mu X^{3}(\tau ),
\end{equation}
where $X(\tau )\equiv x(\tau /\Omega )$ and the dot represents derivation
with respect to $\tau $. Notice that the period of $X(\tau )$ is independent
of $\mu $. On expanding both $X(\tau )$ and $\Omega ^{2}$ in powers of $\mu $%
,
\begin{equation}
\Omega ^{2}=\sum_{n=0}^{\infty }\alpha _{n}\mu ^{n}\;,\qquad X(\tau
)=\sum_{n=0}^{\infty }X_{n}(\tau )\mu ^{n}\;,
\end{equation}
with $\alpha _{0}=1$, one obtains the set of equations
\begin{eqnarray}
\ddot{X}_{n}(\tau )+X_{n}(\tau ) &=&-\theta (n-1)\left[
\sum_{j=0}^{n-1}\sum_{k=0}^{n-1-j}X_{j}(\tau )X_{k}(\tau )X_{n-1-j-k}(\tau
)\right. \   \nonumber \\
&&\left. +\sum_{j=1}^{n}\alpha _{j}\ddot{X}_{n-j}(\tau )\right]  \nonumber \\
n &=&0,1,2,\ldots  \label{eq:LP_PT}
\end{eqnarray}
where $\theta $ is the Heaviside function.

If we choose the coefficients $\alpha _{n}$ in order to remove the resonant
terms, then the solutions $X_{n}(\tau )$ are periodic, have the general form
\begin{equation}
X_{n}(\tau )=\sum_{j=0}^{n}A_{nj}\cos {[(2j+1)\tau ]}\;,  \label{eq:LP_Xn}
\end{equation}
and satisfy the initial conditions
\begin{equation}
X_{n}(0)=\delta _{n0},\,\dot{X}_{n}(0)=0\ .  \label{eq:LP_bc}
\end{equation}

This approach not only gives us a truly periodic approximate solution but
also the frequency of the motion in the form of a power series. For example,
through third order we have:
\begin{equation}
\Omega _{(3)}^{2}=1+\frac{3}{4}\mu -\frac{3}{128}\mu ^{2}+\frac{9}{512}\mu
^{3}\;.
\end{equation}
Since the motion is unbounded for $\mu <-1$ then the $\mu $--power series is
expected to be valid only for $|\mu |<1$. This is a serious limitation of
the LPT because the motion is periodic for all $\mu >-1$ as indicated above.

\subsection{The Lindstedt--Poincar\'{e} Technique and the Linear Delta
Expansion}

In order to improve the LPT, Amore \emph{et al}. \cite{AA03,AL04,AA03b,AM04}
proposed its combination with the linear delta expansion (LDE). The LDE is a
variational perturbation theory like the one proposed some time ago as a
renormalization of the perturbation series in quantum mechanics \cite{K81}
which has proved being suitable for the treatment of a wide variety of problems
\cite{F00,AFC90}.

In order to apply the LDE to the Duffing oscillator we rewrite Eq. (\ref
{eq:Duffing}) as
\begin{equation}
\frac{d^{2}x(t)}{dt^{2}}+\left( 1+\lambda ^{2}\right) x(t)=\delta \left[
-\mu x^{3}(t)+\lambda ^{2}x(t)\right]
\end{equation}
where $\lambda $ is a variational parameter and $\delta $ is an
order--counting parameter. When we set $\delta $ equal to one we recover
the original equation (\ref{eq:Duffing}) that is independent of $\lambda $.
Following the LPT outlined above we change the time variable and obtain
\begin{equation}
\Omega ^{2}\ddot{X}(\tau )+(1+\lambda ^{2})X(\tau )=\delta \left[ -\mu
X^{3}(\tau )+\lambda ^{2}X(\tau )\right] .
\end{equation}
Next we expand both $X$ and $\Omega ^{2}$ in powers of $\delta $ and proceed
exactly as indicated above for the Lindstedt--Poincar\'{e} technique, except
that in this case we choose $\alpha _{0}=1+\lambda ^{2}$. The resulting
perturbation equation of order $n$ reads
\begin{eqnarray}
\ddot{X}_{n}(\tau )+X_{n}(\tau ) &=&-\frac{\theta (n-1)}{\alpha _{0}}\left[
\sum_{j=0}^{n-1}\sum_{k=0}^{n-1-j}X_{j}(\tau )X_{k}(\tau )X_{n-1-j-k}(\tau
)\right. \;  \nonumber \\
&&+\left. \sum_{j=1}^{n}\alpha _{j}\ddot{X}_{n-j}(\tau )-\lambda
^{2}X_{n-1}(\tau )\right] ,  \nonumber \\
n &=&0,1,\ldots
\end{eqnarray}
and the solutions satisfy the initial conditions (\ref{eq:LP_bc}).
As before, the coefficient $\alpha _{n}$ is set to remove the
resonant term from the equation of order $n$. At third order we
find, after setting $\delta =1$,%
\begin{equation}
\Omega _{(3)}^{2}=1+\frac{3\mu }{4}-\frac{3\mu ^{2}}{128(1+\lambda ^{2})}+%
\frac{3\mu ^{2}(3\mu -4\lambda ^{2})}{512(1+\lambda ^{2})^{2}}\;.
\end{equation}
We assume that the optimum value of the arbitrary variational parameter $%
\lambda $ is given by the principle of minimal sensitivity (PMS)~\cite{S81}
\begin{equation}
\frac{d\Omega _{(N)}}{d\lambda }=0\;
\end{equation}
that gives us $\lambda _{PMS}^{(N)}$ and an approximate expression of order $%
N$. At third order we have
\begin{equation}
\lambda _{PMS}^{(3)}=\frac{\sqrt{3\mu }}{2}
\end{equation}
and the approximate frequency results to be
\begin{equation}
\Omega _{(3)}^{2}=\frac{69\mu ^{2}+192\mu +128}{32(3\mu +4)}\;.
\end{equation}
We have already named this approach
Lindstedt--Poincar\'{e}--Linear--Delta--Expansion (LPLDE).

The LPLDE approximants are not $\mu $--power series and prove to be valid
for all $\mu >-1$.

\section{\label{sec:ALP} Alternative Lindstedt--Poincar\'{e} Technique}

There is another approach that resembles the LPT discussed above that
consists of rewriting Eq. (\ref{eq:Duffing}) as \cite{M70}
\begin{equation}
\frac{d^{2}x(t)}{dt^{2}}+\left( \Omega ^{2}-\sum_{n=1}^{\infty }\alpha
_{n}\mu ^{n}\right) x(t)=-\mu x^{3}(t)\;.
\end{equation}
Expanding $x(t)$ in a $\mu $--power series and considering $\Omega $
independent of this parameter, the perturbative equation of order $n$ becomes
\begin{eqnarray}
\frac{d^{2}x_{n}(t)}{dt^{2}}+\Omega ^{2}x_{n}(t) &=&-\theta (n-1)\left[
\sum_{j=0}^{n-1}\sum_{k=0}^{n-1-j}x_{j}(t)x_{k}(t)x_{n-1-j-k}(t)\right. \;
\nonumber \\
&&\left. +\sum_{j=0}^{n}\alpha _{j}x_{n-j}(t)\right] .
\end{eqnarray}
The general form of the solutions $x_{n}(t)$ which fulfill the initial
conditions (\ref{eq:LP_bc}) is
\begin{equation}
x_{n}(t)=\sum_{j=0}^{n}A_{nj}\cos {[(2j+1)\Omega t]}\;,  \label{eq:SoluALP}
\end{equation}
provided that the coefficients $\alpha _{n}$ remove the resonant
terms. These coefficients turn out to depend upon $\Omega $ and,
as a result, any truncated expansion for $\Omega ^{2}$ is a
self--consistent equation for the frequency. For example, solving
for $\Omega _{(3)}^{2}$ in the equation of third order
\begin{equation}
\Omega _{(3)}^{2}=1+\frac{3\mu }{4}-\frac{3\mu ^{2}}{128\Omega _{(3)}^{2}}%
+0\mu ^{3}\;,
\end{equation}
we obtain
\begin{equation}
\Omega _{(3)}^{2}=\frac{\sqrt{30\mu ^{2}+96\mu +64}+2(3\mu +4)}{16}\;.
\end{equation}
We have earlier called this approximation ALPT.

\section{\label{sec:Other_AO} Other Anharmonic Potentials}

In order to compare the approaches outlined above we also consider
anharmonic oscillators with potentials
\begin{equation}
V(x)=\frac{1}{2}x^{2}+\frac{\mu }{2N}x^{2N}\;,\;N=2,3,\ldots
\end{equation}
that lead to the equations of motion
\begin{equation}
\frac{d^{2}x(t)}{dt^{2}}+x(t)=-\mu x^{2N-1}\;,  \label{eq:AnhanPot}
\end{equation}
with exactly the same initial conditions considered before for the Duffing
oscillator, which is a particular case with $N=2$. Notice that there is
periodic motion for all $\mu >-1$ consistent with the initial conditions.

Here we apply the LPLDE and the ALPT to the cases $N=3$ and $N=4$.
The procedure is straightforward, and we only remark that after
removal of resonant terms the general form of the LPT and LPLDE
solutions is
\begin{equation}
X_{n}(\tau )=\sum_{j=0}^{Nn}A_{nj}\cos {[(2j+1)\tau ]}\;,
\end{equation}
whereas for the ALPT we have
\begin{equation}
x_{n}(t)=\sum_{j=0}^{Nn}A_{nj}\cos {[(2j+1)\Omega t]}\;.
\end{equation}

\section{\label{sec:VDP} The Van der Pol equation}

In order to test the performance of the approaches outlined above we also
choose the well--known Van der Pol (VdP) equation
\begin{equation}
\frac{d^{2}x(t)}{dt^{2}}+x(t)=\mu \ [1-x^{2}(t)]\ \frac{dx(t)}{dt}.
\label{eq:vdp}
\end{equation}
because the behaviour of its solutions is completely different from those of
the anharmonic oscillators discussed above.

The VdP equation exhibits a limit cycle and leads to oscillations with a
definite period that depends on the coupling $\mu $. However, unlike the
anharmonic oscillators the VdP equation does not correspond to a
conservative system as the driving term either damps or enhances the
oscillation depending upon the size of $x$. This equation has already been
treated by means of the LPLDE approach ~\cite{AM04}; here we apply and
compare it with the ALP method.

\subsection{The LPLDE Technique}

We first rewrite the VdP equation (\ref{eq:vdp}) as \cite{AM04}
\begin{equation}
\Omega ^{2}\ddot{x}(\tau )+(1+\lambda ^{2})x(\tau )=\delta \{\mu \Omega
[1-x^{2}(\tau )]\dot{x}(\tau )+\lambda ^{2}x(\tau )\}\;.
\end{equation}
In this case we expand $\Omega $ (instead of $\Omega ^{2}$) in powers of $%
\delta $
\begin{equation}
\Omega =\sum_{n=0}^{\infty }\alpha _{n}\delta ^{n}\ \ \ ,\ \ \
x=\sum_{n=0}^{\infty }x_{n}(\tau )\delta ^{n}\;,  \label{eq:exp_vdp}
\end{equation}
and choose $\alpha _{0}=1+\lambda ^{2}$. We thus obtain
\begin{eqnarray}
\ddot{x}_{n}(\tau )+x_{n}(\tau ) &=&\frac{\theta (n-1)}{\alpha _{0}}\left[
\mu \sum_{j=0}^{n-1}\alpha _{j}\dot{x}_{n-1-j}-\sum_{j=1}^{n}\alpha _{j}\ddot{x}_{n-j}(\tau )\right.  \nonumber \\
&&\left. -\mu \sum_{i=0}^{n-1}\sum_{j=0}^{n-1-i}\sum_{k=0}^{n-1-i-j}\alpha
_{i}x_{j}(\tau )x_{k}(\tau )\dot{x}_{n-1-i-j-k}(\tau )\right] .
\end{eqnarray}
The general form of the solutions is
\begin{equation}
x_{n}(\tau )=\sum_{j=0}^{n}\left[ c_{nj}\cos (2j+1)\tau +d_{nj}\sin
(2j+1)\tau \right] \ ,
\end{equation}
that satisfy the initial conditions
\begin{equation}
x_{n}(0)=A_{n}\ \ \ ,\ \ \ \dot{x}_{n}(0)=0  \label{eq:bcvdp}
\end{equation}
for $n\geq 1$. The appropriate choice of $\alpha _{n}$ and
$A_{n-1}$ enables us to remove all resonant terms at order $n$.

\subsection{The ALPT}

Proceeding as in the case of the anharmonic oscillators we substitute the
expansion for the frequency $\Omega ^{2}$ into the VdP equation and obtain
\begin{eqnarray}
\frac{d^{2}x_{n}(t)}{dt^{2}}+\Omega ^{2}x_{n}(t) &=&\theta (n-1)\left[
\sum_{j=1}^{n}\alpha _{j}x_{n-j}(t)+\frac{dx_{n-1}(t)}{dt}\right.  \nonumber
\\
&&\left. -\sum_{j_{1}=0}^{n-1}\sum_{j_{2}=0}^{n-1-j_{1}}x_{j_{1}}(t)\
x_{j_{2}}(t)\ \frac{dx_{n-1-j_{1}-j_{2}}(t)}{dt}\right]
\end{eqnarray}
The general form of the solutions in this case is
\begin{equation}
x_{n}(t)=\sum_{j=0}^{n}\left\{ c_{nj}\cos [(2j+1)\Omega t]+d_{nj}\sin
[(2j+1)\Omega t]\right\} \ ,  \label{eq_x}
\end{equation}
that satisfy the initial conditions (\ref{eq:bcvdp}).

The appropriate choice of $\alpha _{n}$ and $A_{n-1}$ enable us to remove
secular terms at order $n$. As before, the coefficients $\alpha _{n}$ of the
expansion of $\Omega ^{2}$ in powers of $\mu $ depend on $\Omega $ that one
obtains as a real solution of the partial sum at a given order.

\section{\label{sec:Results} Results and Discussion}

\begin{figure}[tbp]
\begin{center}
\includegraphics[width=9cm]{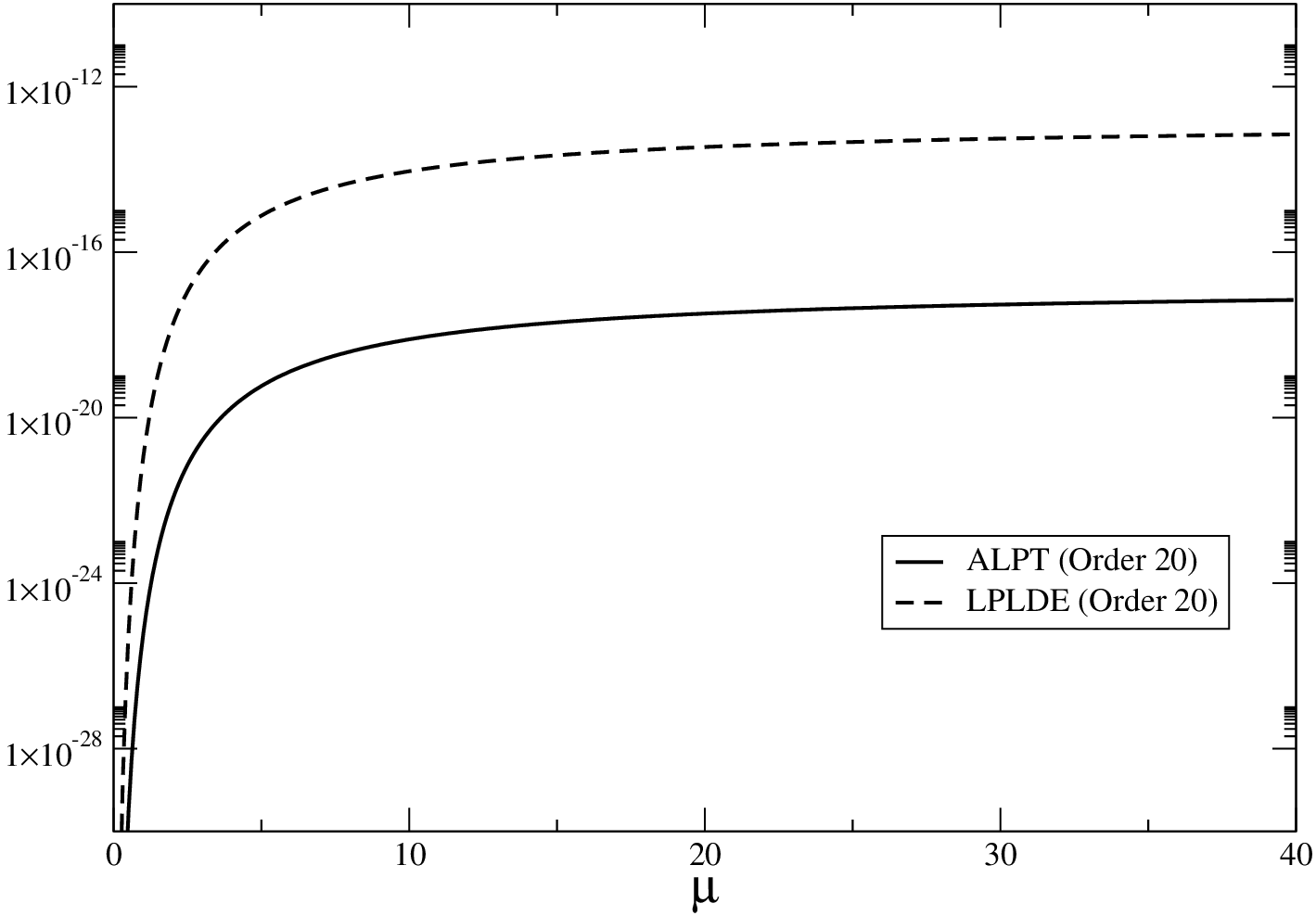}
\end{center}
\caption{Error over the frequency as a function of $\mu$. }
\label{fig1}
\end{figure}

\begin{figure}[tbp]
\begin{center}
\includegraphics[width=9cm]{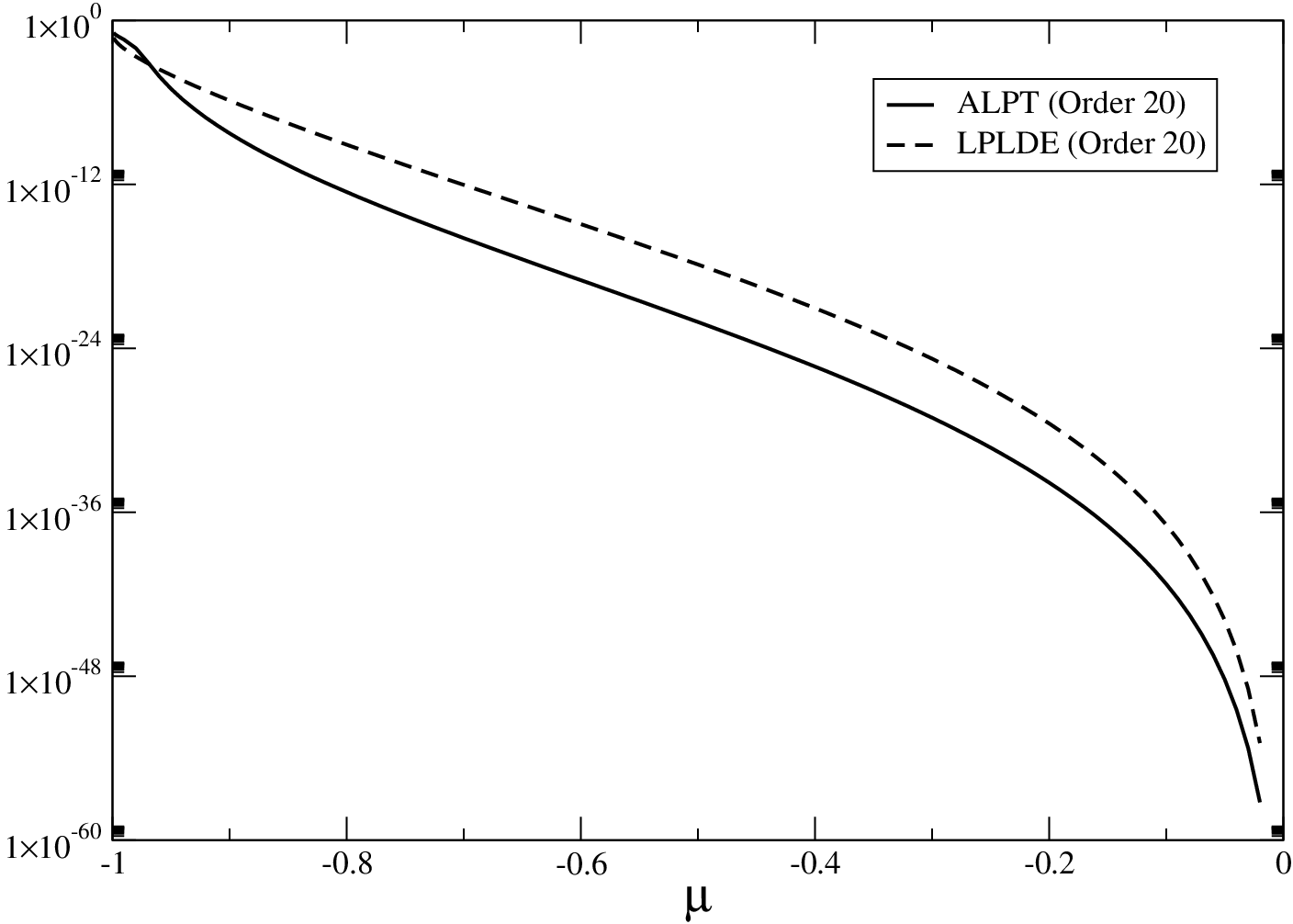}
\end{center}
\caption{Error over the frequency as a function of $\mu$, for $-1 \leq \mu \leq 0$. }
\label{fig1a}
\end{figure}

We first consider the results of the LPLDE and ALPT for the Duffing
oscillator. In the case of the LPLDE we use at all orders the variational
parameter $\lambda _{PMS}$ calculated at third order. Fig.~(\ref{fig1}) shows
the error over the frequency, defined as $\Delta \equiv |\Omega_{exact}-\Omega_{approx}|$,
as a function of $\mu $ ($\mu \geq 0$) produced by partial sums through order $20$ for both approaches.
We observe that the LPLDE technique, which yields much more accurate results than the straightforward
LPT, is less accurate than the ALPT. A similar behavior can be observed in Fig.~(\ref{fig1a}), where
the error over the frequency is plotted for $-1 \leq \mu \leq 0$. Interestingly, close to
$\mu=-1$, which corresponds to a condition of (unstable) equilibrium, the LPLDE method
performs better than the ALPT.

\begin{figure}[tbp]
\begin{center}
\includegraphics[width=9cm]{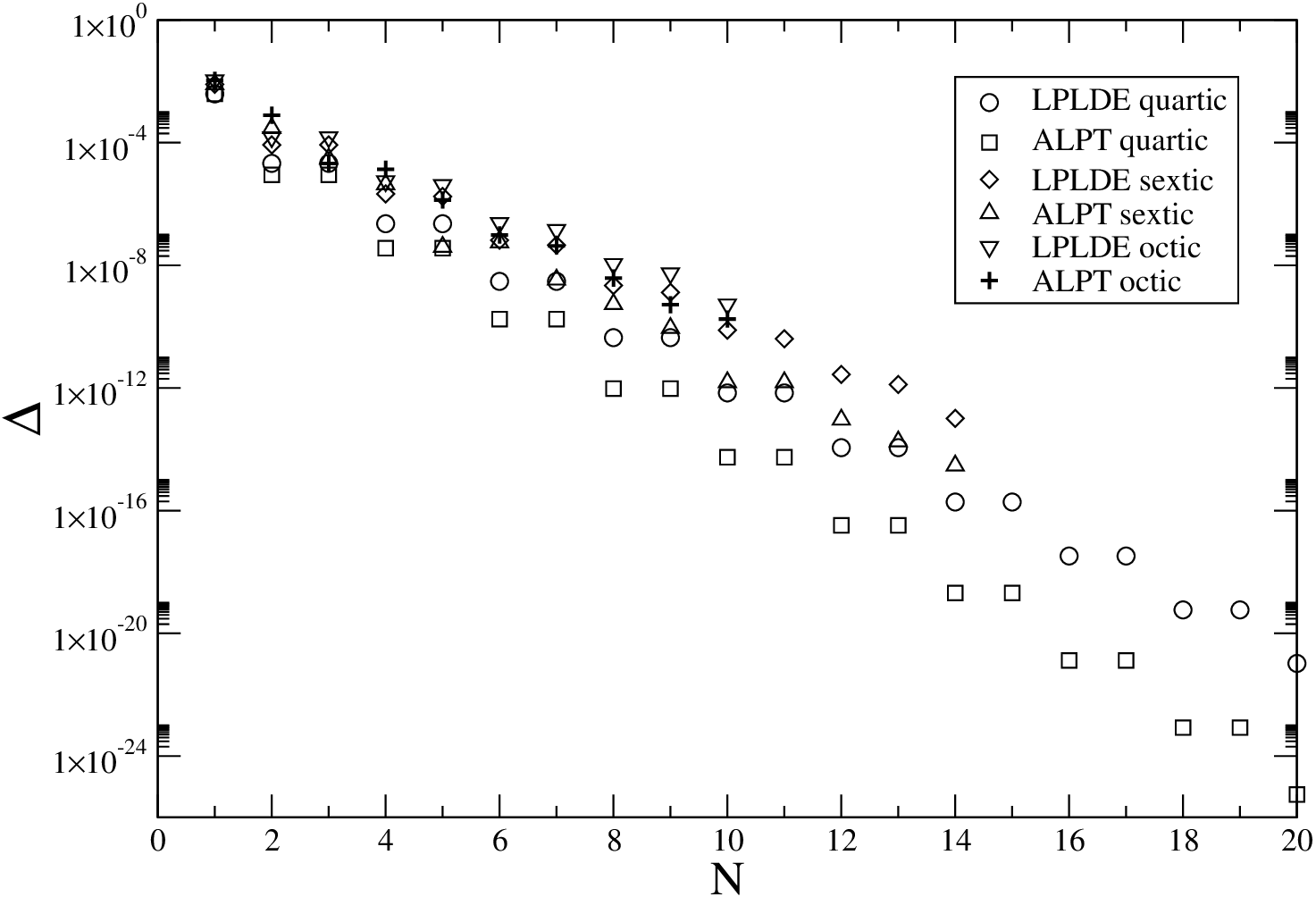}
\end{center}
\caption{Error over the frequency as a function of the number of terms
taking $\mu=1$ for different anharmonic potentials. }
\label{fig2}
\end{figure}

\begin{figure}[tbp]
\begin{center}
\includegraphics[width=9cm]{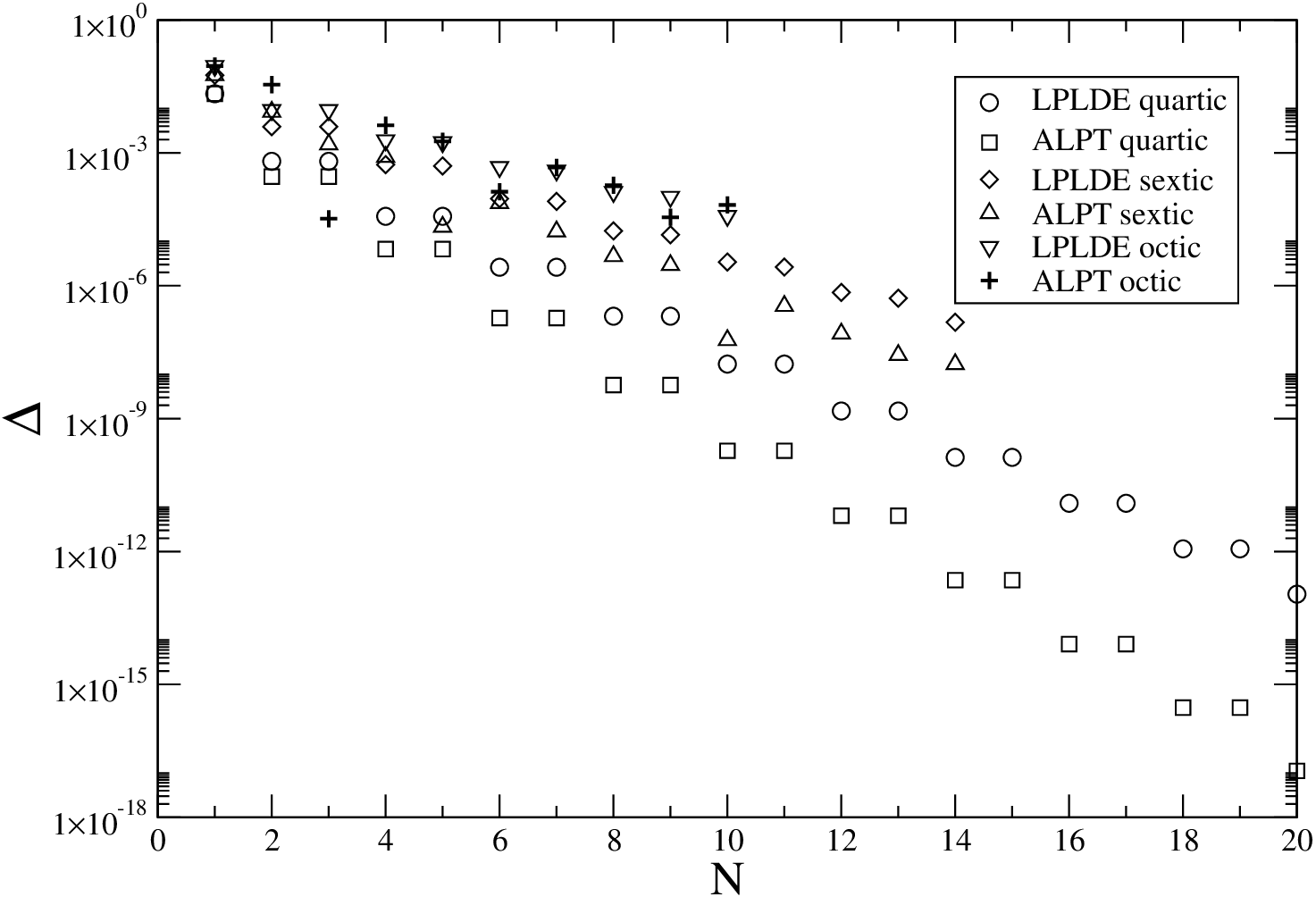}
\end{center}
\caption{Error over the frequency as a function of the number of terms
taking $\mu=100$ for different anharmonic potentials. }
\label{fig3}
\end{figure}

Similar conclusions can be drawn in the case of anharmonic oscillators with greater $N$.
Figs. (\ref{fig2}) and (\ref{fig3}) show the error over the frequency as a function of the order of
approximation for quartic (Duffing), sextic and octic anharmonic oscillators
with $\mu =1$ and $\mu =100$, respectively.

\begin{figure}[tbp]
\begin{center}
\includegraphics[width=9cm]{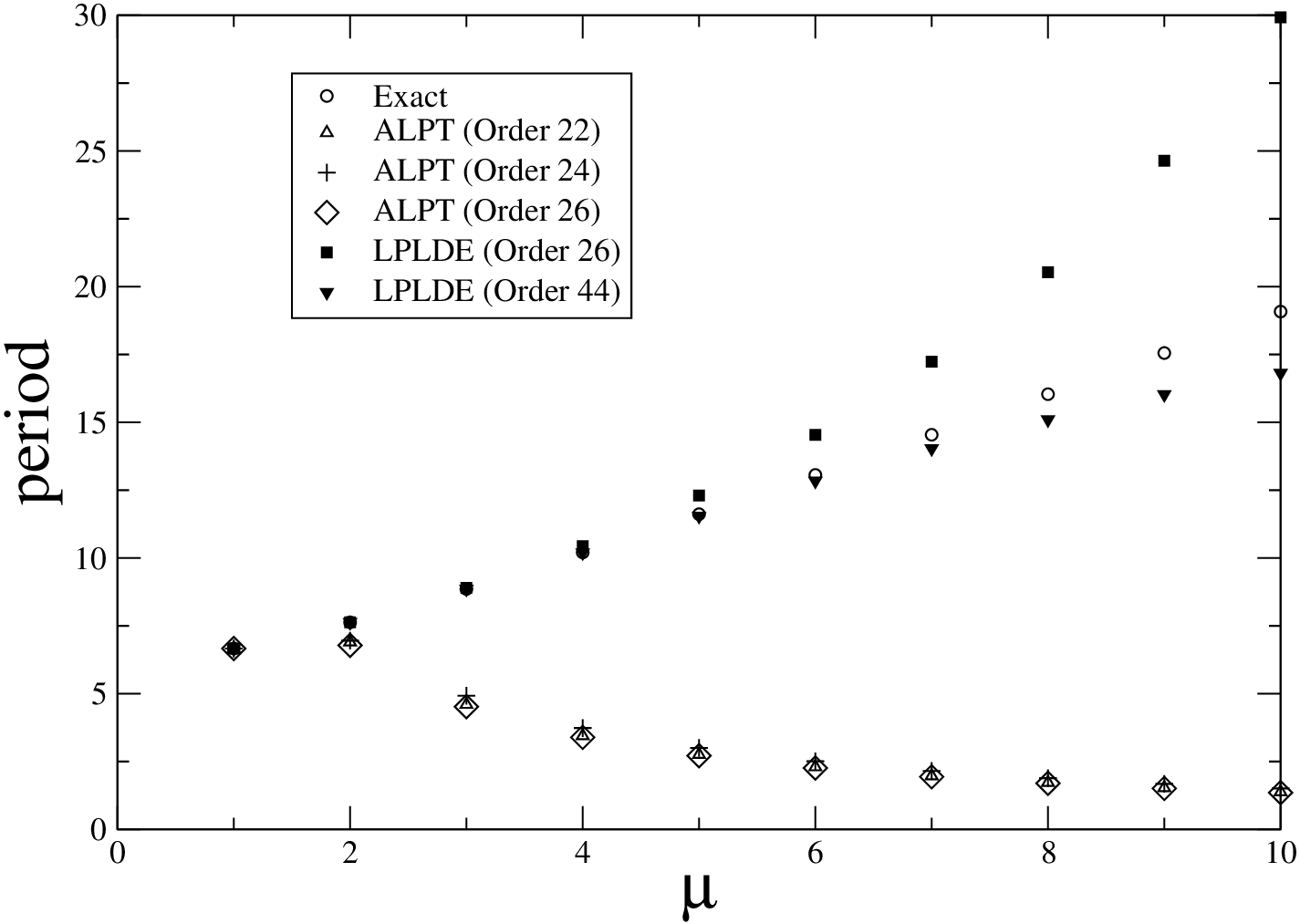}
\end{center}
\caption{Period of the Van der Pol oscillator as a function of $\mu$.}
\label{fig4}
\end{figure}

The situation is different in the case of the nonconservative VdP equation
because the ALPT exhibits a great convergence rate, but it yields a wrong
frequency for $\mu >\simeq 2$, which suggests a wrong dependence on this
coupling parameter. On the other hand, the LPLDE yields reasonable results
for any value of $\mu $, as shown in Fig.~(\ref{fig4}).

Our results suggest that the ALPT method is more accurate for conservative
systems but fails for nonconservative ones, while, on the other hand, the
LPLDE yields reasonable results in both cases. At present we are unable to
explain the failure of the ALPT for nonconservative systems and we believe
that this subject deserves further investigation.

{\bf Acknowledgements} P.A. acknowledges support of Conacyt grant no. C01-40633/A-1. P.A. and A.R. acknowledge
support of the Alvarez-Buylla fund of the University of Colima.


\end{document}